\documentclass[useAMS,usenatbib]{mnras}
\usepackage{rotating}
\usepackage{booktabs}
\usepackage{setspace}
\usepackage{txfonts}
\usepackage{float}
\usepackage{amssymb}
\usepackage{epsfig}
\usepackage{color}
\usepackage{graphicx}
\usepackage{multirow}
\usepackage{longtable}
\usepackage{verbatim}
\usepackage[flushleft]{threeparttable}
\usepackage{array}
\usepackage{natbib}

\newcolumntype{L}[1]{>{\raggedright\let\newline\\\arraybackslash\hspace{0pt}}m{#1}}
\newcolumntype{C}[1]{>{\centering\let\newline\\\arraybackslash\hspace{0pt}}m{#1}}
\newcolumntype{R}[1]{>{\raggedleft\let\newline\\\arraybackslash\hspace{0pt}}m{#1}}

\title[The {\it XMM-Newton} spectra of the 2012 outburst of 4U~1630$-$47 revisited]
{
\begin{center}
The {\it XMM-Newton} spectra of the 2012 outburst of the black-hole candidate 4U~1630$-$47 revisited
\end{center}
}

\author[Yanan Wang et al.]{Yanan Wang$^{1}$\thanks{E-mail: yanan@astro.rug.nl}, Mariano M\'endez$^{1}$\\
$^{1}$Kapteyn Astronomical Institute, University of Groningen, PO BOX 800, NL-9700 AV Groningen, the Netherlands\\}

\begin{document}

\date{Accepted ? December ?. Received ? December ?; in original form ? December ?}

\maketitle

\label{firstpage}

\begin{abstract}
Recent {\it XMM-Newton} observations of the black-hole candidate 4U~1630$-$47 during the 2012 outburst 
revealed three relativistically Doppler-shifted emission lines that were interpreted as arising from baryonic matter 
in the jet of this source. Here we reanalyse those data and find an alternative model that, with less free parameters 
than the model with Doppler-shifted emission lines, fits the data well. In our model we allow the 
abundance of S and Fe in the interstellar material along the line of sight to the source to be non solar. Among 
other things, this significantly impacts the emission predicted by the model at around 7.1 keV, where the edge of 
neutral Fe appears, and renders the lines unnecessary. The fits to all the 2012 {\it XMM-Newton} observations of 
this source require a moderately 
broad emission line at around 7 keV plus several absorption lines and edges due to highly ionised Fe and Ni, which
reveal the presence of a highly-ionised absorber close to the source. Finally, our model also fits well the observations 
in which the lines were detected when we apply the most recent calibration files, whereas the model with the three Doppler-shifted emission lines 
does not.
\end{abstract}

\begin{keywords}
accretion, accretion disk--binaries: 
X-rays: individual (4U~1630$-$47)

\end{keywords}

\section{INTRODUCTION}
The soft X-ray transient 4U~1630$-$47 \citep{Jones1976, Parmar1995} shows regular outbursts every 600-690 days \citep{Abe2005,Tomsick2014}. 
The source has been classified as a black hole \citep{Parmar1986} because of the similarity of 
its spectral and timing properties to those of systems with measured black-hole masses \citep[e.g.,][]{Barret1996,Abe2005}. 
4U~1630$-$47 shows strong absorption by neutral material along the line of sight, with a hydrogen column density
$N\rm{_{H}=5-12 \times 10^{22}~cm^{-2}}$ \citep[e.g.,][]{Tomsick1998}, and both IR \citep{Augusteijn2001} and radio emission 
\citep{Hjellming1999} were detected during the 1998 outburst of this source. 
The optical counterpart of 4U~1630$-$47 has not
been identified, mostly due to the high reddening and the location of the source in a crowded star field \citep{Parmar1986}.

Absorption lines due to highly ionised material have been observed in the spectrum of 4U~1630$-$47 
\citep{Kubota2007, Rska2014, DT2014, Miller2015}. Using {\it Suzaku} observations carried out in 2006, 
\cite{Kubota2007} studied these absorption line features in relation to the accretion-disc parameters,
and concluded that the lines were due to a wind. 
Using the same {\it Suzaku} data, \cite{Rska2014} proposed that the absorption lines could be produced 
effectively in the accretion disc atmosphere. 
Using {\it XMM-Newton} observaitons, \cite{DT2014} found a thermally/radiatively
driven disc wind in 4U~1636$-$47; the wind becomes more photoionised as the luminosity of the source increases. 
Recently, \cite{Miller2015} analyzed {\it Chandra} observations of 4U~1630$-$47 and three other galactic black hole candidates.
For 4U~1630$-$47, they found that the wind consists of at least two absorption zones with velocities of $~\rm{-200~km~s^{-1}}$ 
and $~\rm{-2000~km~s^{-1}}$, respectively.
They also found that, in some respects, these zones correspond to the broad-line region in active galactic nuclei.

\cite{DT2013} analysed two {\it XMM-Newton} and two quasi-simultaneous observations with 
the Australia Telescope Compact Array (ATCA) carried out during the 2012 outburst of 4U~1630$-$47.
\cite{DT2013} found three relatively narrow emission lines in the X-ray spectrum of one of these observations
that they identified as arising from baryonic matter in a jet.
The three lines had energies of 4.04~keV, 7.28~keV and 8.14~keV, respectively, which \cite{DT2013}
interpreted as the red- and blueshifted component of Fe~{\sc xxvi} Ly$\alpha$
and the blueshifted component of Ni~{\sc xxviii} Ly$\alpha$, respectively.
From the radio data, \cite{DT2013} confirmed that there was an optically thin jet in 4U~1630$-$47 at the time of that observation. 
\cite{Hori2014} investigated {\it Suzaku} and Infrared Survey Facility observations of 4U~1630$-$47 during the same outburst,
at a time when the source was in the very high state. These observations were carried out three to five days after the {\it XMM-Newton} observations of \cite{DT2013}.
The {\it Suzaku} X-ray spectra, however, did not show the Doppler-shifted emission lines of the jet reported by \cite{DT2013}.  
Using {\it Chandra} and ATCA observations taken 
eight months prior to the {\it XMM-Newton} observations of \cite{DT2013},                                              
\cite{Neilsen2014} reported a similar result to that of \cite{Hori2014}.

\begin{table*}
\caption{\label{tab:1}{\it XMM-Newton} Observations of 4U 1630$-$47 used in this paper}
\renewcommand{\arraystretch}{1.3}
\footnotesize
\begin{tabular}{lcccc}
\hline \hline
\multirow{2}{*}{ObsID}&Observation Time (UTC)&\multirow{2}{*}{Observation mode}&\multirow{2}{*}{RAWX source}&\multirow{2}{*}{RAWX back}\\
      & (day/month/year hh:mm)&  & & \\
\hline
0670671501-1 & 04/03/2012 11:24 - 04/03/2012 12:27 & Timing &[27,46] &[4,10]\\
0670671501-2 & 04/03/2012 13:43 - 05/03/2012 09:23& Timing& [28,45]&[4,10]\\
0670671301 & 20/03/2012 19:54 - 21/03/2012 02:30 & Timing & [28,45] &[4,10] \\
0670672901 & 25/03/2012  04:14 - 25/03/2012 21:56 & Timing&[28,45] &[4,10]\\
0670673001 & 09/09/2012 21:14 - 10/09/2012 07:49 & Timing& [28,45]&[4,10]\\
0670673101 & 11/09/2012 20:56 - 12/09/2012 05:38 & Burst &[20,51] &[4,10]\\
0670673201 & 28/09/2012 07:16 - 28/09/2012 21:48 & Burst& [20,51]& [4,10]\\ 
\hline
\end{tabular}
\begin{flushleft}
{\bf Notes.} ObsID 0670671501 contains two separate event files in timing mode that
we extracted and fitted separately. We called them 0670671501-1
and 0670671501-2, respectively. RAWX source and RAWX back indicate
the extraction region in the CCD for the source and the background, respectively.
\end{flushleft}
\end{table*}

When fitting the {\it Chandra} data of 4U~1630$-$47, \cite{Neilsen2014} allowed the abundance of Si, S and Ni in the 
component that they fitted to the interstellar absorption to be different from solar but, unfortunately, they do not report the
best fitting values of these parameters. On the other hand, using the Reflection Grating Spectrometer on board {\it XMM-Newton}, 
\cite{Pinto2013} measured the abundance of O, Ne, Mg, and Fe in the interstellar medium (ISM) in the 
direction of nine low-mass X-ray binaries, not including 4U~1630$-$47. Interestingly, they found that the Fe abundance in the neutral ISM
in the direction of these sources ranges between less than 0.02 and 0.50 times the solar abundance. Because the putative lines reported by 
\cite{DT2013} are close to the K-$\alpha$ edges of (neutral) Ca~{\sc i} (4.04 keV), Fe~{\sc i} (7.12 keV), and Ni~{\sc i} (8.34 keV), 
and the column density toward 4U~1630$-$47 is quite high (see above), the results of \cite{Pinto2013} suggest the possibility 
that the emission lines reported by \cite{DT2013} could in fact be an artefact of the model if the incorrect elemental 
abundance in the ISM is used in the fits. \cite[][assumed solar abundance in their fits.]{DT2013}

In this paper, we used the same {\it XMM-Newton} data of 4U~1630$-$47 as \cite{DT2013}, but we explore an alternative model in which 
we allow the abundance of the ISM to be different from solar. We can fit the data well with a model that does not require any 
Doppler-shifted emission lines; instead, our fits yield non-solar abundances of S and Fe in the ISM along the line of sight 
to the source. Our model not only fits the two observations in \cite{DT2013}, but also the other {\it XMM-Newton} observations 
during the 2012 outburst \citep{DT2014}, in which four absorption lines, that we identify as being produced by Fe~{\sc xxv}, 
Fe~{\sc xxvi} and Ni~{\sc xxviii} (or Fe~{\sc xxv}~Ly$\beta$), 
and two absorption K-edges, due to Fe~{\sc xxv} and Fe~{\sc xxvi}, are detected.  Furthermore, the putative Doppler-shifted 
emission lines are not required either using the same model as \cite{DT2013} when we apply the new calibration files to those observations.

\section{OBSERVATIONS AND DATA REDUCTION}

The X-ray data that we used here consist of six observations of 4U~1630$-$47 with {\it XMM-Newton} \citep{Jansen2001}
taken between March 4 and September 28 2012.
We report the details of the observations in Table~\ref{tab:1}. 
We only used data from the European Photon Imaging Camera, EPIC-pn \citep{Strder2001}, which was operated in either burst or timing mode.
To reduce and analyse the raw data we used version 14.0.0 of the {\it XMM-Newton} Scientific Analysis Software (SAS) package 
following standard procedures. 

We used the command {\it epproc} to calibrate the timing- and burst-mode photon event files. 
Following the recommendations of the {\it XMM-Newton} team, for the burst-mode data we also ran the command {\it epfast}. 
This command applies a correction to the energy scale due to charge transfer inefficiency in the CCD in burst mode. 
While there is some discussion in the literature regarding the applicability of this correction \citep[see][and the XMM-Newton Calibration Technical Note of November 2014\footnote{\url{http://xmm.vilspa.esa.es/docs/documents/CAL-TN-0083.pdf}}]{Walton2012}, 
\cite{DT2013} applied this correction during their analysis and therefore, in order to compare to their results, we apply it here. For completeness,
we also reduced the burst-mode observations without applying the {\it epfast} correction.

We selected calibrated events with PATTERN$\leq$4 in the central CCD of the EPIC-pn camera to get the spectrum of 
the source and we extracted a background spectrum from the outer columns of the central CCD (see Table~\ref{tab:1} for the parameters of the extraction regions).

The difference between timing and burst mode in the process of extracting the data is that the spectra of the latter 
are influenced by the value of RAWY, i.e., the CCD row number \citep{Kirsch2006}. 
Following the recommendations of the {\it XMM-Newton} team,
we excluded events with RAWY $>$ 140.
We rebinned the average EPIC-pn spectra before fitting in order to have a minimum of 25 counts in each bin. 
We created the redistribution matrix file (RMF)
and the ancillary response file (ARF) using the SAS tasks {\it rmfgen} and {\it arfgen}, respectively.
Following \cite{DT2013}, we fitted the EPIC-pn spectra between 2 and 10~keV.

We used the spectral analysis package XSPEC v12.8.2 to fit the data, adding a 1\% systematic error
to the model to account for calibration uncertainties.
The models that we used in this paper include
a component to account for photoelectric absorption of the interstellar material along the line of sight.
For this component we used either {\sc tbabs} or {\sc vphabs}; the latter allows variable abundances 
in the interstellar material.
For the emission component we used {\sc diskbb}, a multi-colour disc blackbody \citep{Mitsuda1984},
{\sc powerlaw}, a simple power law, and {\sc gauss}, to account for possible Gaussian emission lines.
We added several Gaussian absorption lines and edges ({\sc edge}), when necessary.
Throughout the paper, we give the 1$\sigma$ errors for all fitted parameters and, when required, the 95\% confidence upper limits.

\begin{table*}
\caption{\label{tab:2}Best-fitting parameters for the two burst-mode observations of 4U~1630$-$47 based on the old calibration using two models}
\renewcommand{\arraystretch}{1.3}
\setlength{\tabcolsep}{8pt}
\scriptsize
\begin{tabular}{llcccc}
\hline \hline

&&\multicolumn{2}{c}{Model of \cite{DT2013}}& \multicolumn{2}{c}{Our model} \\
ObsID & & 0670673101 & 0670673201 & 0670673101 & 0670673201 \\

\hline
 {\sc tbabs/vhpabs} & $N_{\rm H}~(10^{22}~\mathrm{cm}^{-2})$ & $8.80\pm0.05$ & $8.85\pm0.08$ & $15.0\pm0.2~^{[1]}$&$15.0\pm0.2~^{[1]}$\\
 & S/S$_{\odot}$     & $1.0^f$ & $1.0^f$ & $1.32\pm0.06~^{[2]}$ & $1.32\pm0.06~^{[2]}$\\ 
 & Fe/Fe$_{\odot}$  & $1.0^f$ & $1.0^f$ & $0.54\pm0.07 ~^{[3]}$ & $0.54\pm0.07 ~^{[3]}$\\
\hline
\multirow{2}{*}
{\sc diskbb} & $kT_{in}$ (keV) & $1.73\pm0.01$ & $1.75\pm0.02$ &$1.67\pm0.01$ & $1.68\pm0.02$ \\
 &$k_{dbb}$ & $90.7\pm2.0$ & $91.9\pm3.6$ &$107.7\pm3.0$ & $110.6\pm3.8$\\
\hline
\multirow{2}{*}
{\sc powerlaw} & $\Gamma$ & $2^{f}$ & $2^{f}$ & $2^f$ & $2^f$\\
&$k_{pow}$ & $<0.23$ & $1.03\pm0.14$ & $<0.44$ & $1.10\pm0.09$\\ 
\hline
\multirow{4}{*}
{\sc gauss$_1$} & $E$ (keV) & $4.02\pm0.06~^{[4]}$ & $4.02\pm0.06~^{[4]}$ & $7.0_{-0.05}^{+0p}~^{[5]}$ & $7.0_{-0.05}^{+0p}~^{[5]}$\\
 & $\sigma$ (eV) & $165_{-53}^{+47}~^{[6]}$ & $165_{-53}^{+47}~^{[6]}$ & $183_{-79}^{+108}~^{[7]}$ & $183_{-79}^{+108}~^{[7]}$\\
 & $k_{gau}$ & $<2.3$ & $5.8\pm1.8$ & $<2.3$ & $1.61\pm0.6$\\
 & $W$ (eV) & $<15.9$ & $13.8\pm3.6$ & $<12.2$ & $15.4\pm6.9$\\
\hline
\multirow{4}{*}
{\sc gauss$_2$} & $E$ (keV) & $7.24\pm0.04~^{[8]}$ & $7.24\pm0.04~^{[8]}$ & & \\
&$\sigma$ (eV) & $165_{-53}^{+47}~^{[6]}$ & $165_{-53}^{+47}~^{[6]}$ & & \\
&$k_{gau}$ & $<0.9$ & $3.0\pm0.7$ & & \\
&$W$ (eV) & $<21.7$ & $31.2\pm6.7$ & & \\
\hline
\multirow{4}{*}{{\sc gauss$_3$}}&$E$ (keV) & $8.12\pm0.10~^{[9]}$ & $8.12\pm0.10~^{[9]}$ & & \\
&$\sigma$ (eV) & $165_{-53}^{+47}~^{[6]}$ & $165_{-53}^{+47}~^{[6]}$ & & \\
&$k_{gau}$ & $<0.5$ & $1.5\pm0.6$ & & \\
&$W$ (eV) & $<42.8$ & $23.7_{-8.0}^{+7.5}$ & & \\
\hline
$\chi^{2}_{\nu}$&&\multicolumn{2}{c}{0.98 for 256 d.o.f.}& \multicolumn{2}{c}{0.86 for 259 d.o.f.}\\
\hline
\end{tabular}
\begin{flushleft}
\begin{tablenotes}
$N_{H}$ is the column density of the neutral absorber.\\
S/S$_\odot$ and Fe/Fe$_\odot$ are, respectively, the sulphur and iron abundances, in solar units, of the absorber along the line of sight.\\
$k_{dbb}$, equal to the cosine of the inclination of the accretion disc with respect to the line of sight times the square of the ratio of the inner radius of the disc in km and the distance to the source in units of 10 kpc, $k_{pow}$, in units of $\rm photons~keV^{-1}cm^{-2}s^{-1}$ at 1~keV, and $k_{gau}$, in units of 10$\rm^{-3}~photons~cm^{-2}s^{-1}$, are, respectively, the normalisation of the {\sc diskbb}, {\sc powerlaw} and {\sc gauss} components.\\
$W$ is the equivalent width of the line.\\
Parameters with the same number in between square brackets were linked to be the same during the fit.\\
$^{f}$ This parameter was kept fixed at the given value during the fits.\\
$^{p}$ The energy of this emission line pegged at the upper limit. 
\end{tablenotes}
\end{flushleft}
\end{table*}

\section{RESULTS}
To compare our results with those of \cite{DT2013}, we first fitted the two burst-mode observations, separately from the timing-mode observations, using the
same calibration files that \cite{DT2013} used. We then fitted our model simultaneously to the two 
burst- and the four timing-mode observations. Finally we fitted the same model only to the
burst-mode observations using the latest calibration.

\begin{figure}
\centering
\includegraphics[width=7cm, angle=270]{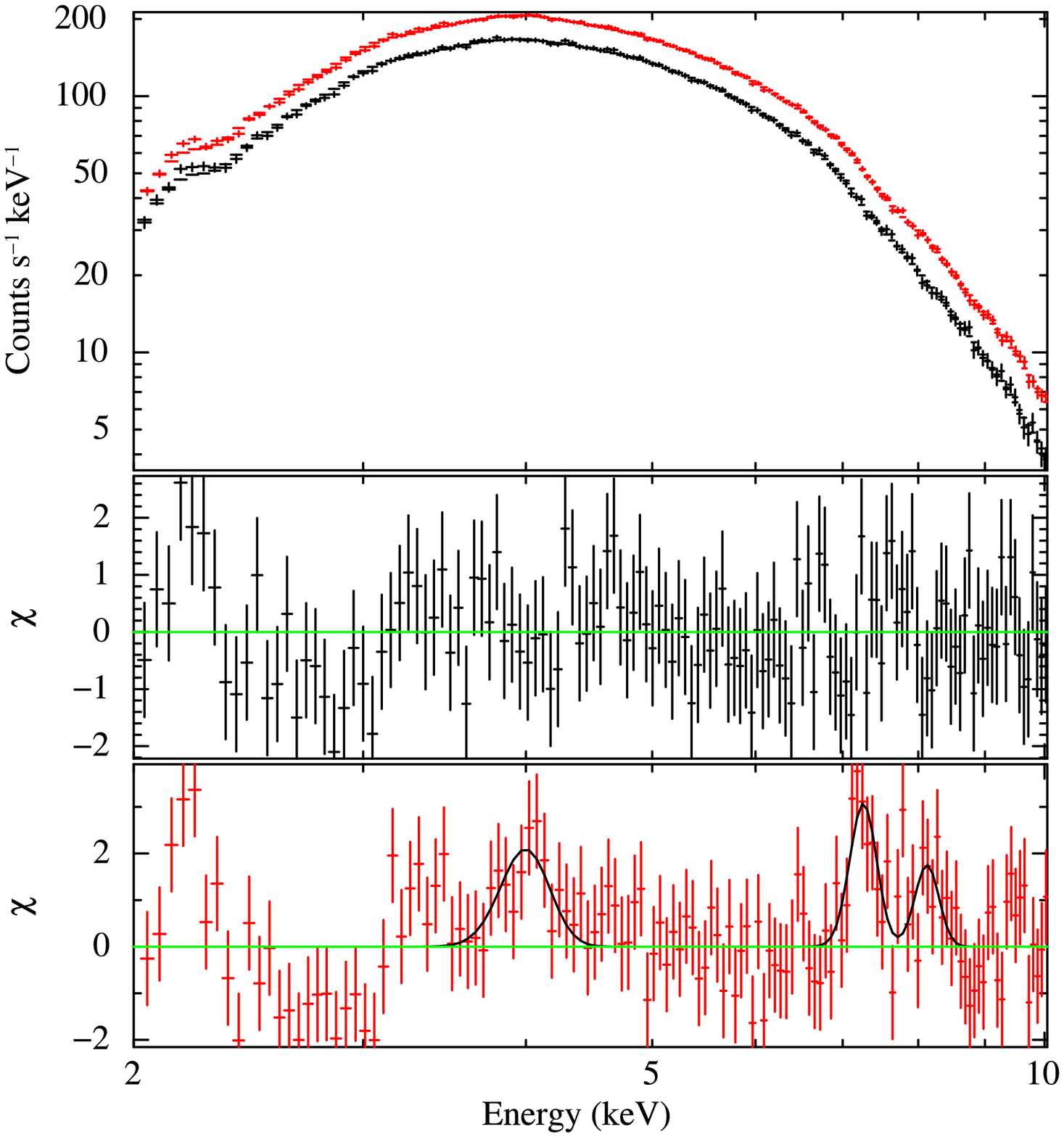}
\caption{\label{fig:1}X-ray spectra of ObsID 0670673101 and 0670673201 of 4U 1630$-$47 fitted with the model of \protect\cite{DT2013}.
The second panel is the residual of the best-fitting of ObsID 0670673101; the third panel is the residual of 
ObsID 0670673201 when the strength of the three {\sc gauss} components is set to zero. For this, and the other figures, the residuals are the data minus the model divided by the error.}.
\end{figure}

\begin{figure}
\centering
\includegraphics[width=7cm, angle=270]{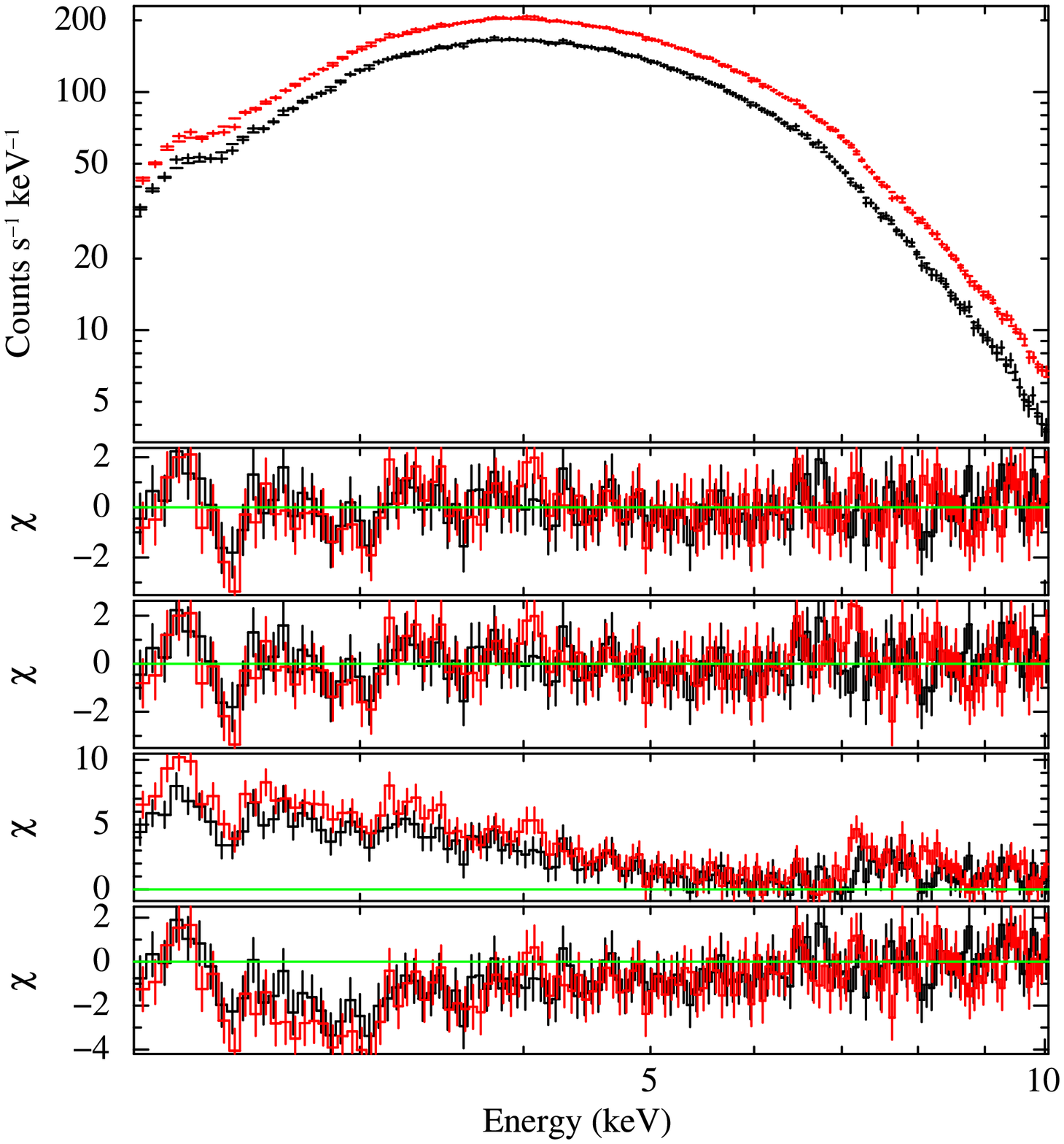}
\caption{\label{fig:2}X-ray spectra of ObsIDs 0670673101 and 0670673201 of 4U 1630$-$47 fitted simultaneously with the alternative model that we proposed here.
The second panel shows the residual of the best-fitting model to the two observations; the following panels are
the residuals of best-fitting model when the strength of the {\sc gauss} component is set to zero (third panel from the top),
the abundance of S in {\sc vphabs} is set to solar (fourth panel from the top), and the abundance of Fe
in {\sc vphabs} is set to solar (bottom panel).}
\end{figure}

\subsection{Fits to the two burst-mode observations using the old calibration files}
Following \cite{DT2013}, 
we first used the model {\sc tbabs}*({\sc diskbb}+{\sc powerlaw}+{\sc gauss$_1$}+{\sc gauss$_2$}+{\sc gauss$_3$}) 
to fit the two burst-mode observations. To reproduce their procedures as close as possible, we allowed $N\rm{_{H}}$ to vary between the two observations.  Similar to \cite{DT2013}, we found three emission lines in ObsID  0670673201, 
but not in ObsID 0670673101. In the latter case we calculated the upper limit to those lines assuming that they had the same energy and width as the lines in the other observation.
We got values of the parameters that, except for $N\rm{_{H}}$ and the normalisation of the {\sc diskbb} and the {\sc gauss} components, were similar to those in \cite{DT2013}. The difference in the normalisations is likely due to the fact that we considered the background spectrum in the analysis, whereas \cite{DT2013} did not. 
We give the best-fitting parameters for this model in Table~\ref{tab:2}, and we plot 
the X-ray spectra and best-fitting model of the two burst-mode observations in Figure~\ref{fig:1}.
As in \cite{DT2013}, to highlight the three emission lines we set their normalisations to zero in the residuals plot.

We then fitted an alternative model to the same data, in which we replaced the {\sc tbabs} component by
the {\sc vphabs} component, and we kept only one of the {\sc gauss} emission components. The other components were the same as those in the model of \cite{DT2013}. In this case we fitted the same model to both spectra simultaneously and, since the interstellar absorption 
along the line of sight to the source should not change, we linked the parameters of {\sc vphabs} between the 
two observations. 
We changed the default solar abundances model ({\sc abund} in XSPEC) to the abundances of \cite{Wilms2000} and the default 
photoelectric absorption cross-sections table ({\sc xsect} in XSPEC) to that given by \cite{Verner1996}.
We fixed the photon index, $\Gamma$, of the {\sc powerlaw} component at 2, since it could not be well constrained in the fits.

One by one, we let the abundance of C, N, O, Ne, Mg, Si, S, Ca, Fe, and Ni in {\sc vphabs} free to fit the data, while the other element abundances were kept fixed at the solar values. Except for the case of S and Fe, the best-fitting abundances were consistent with solar, and hence we eventually left the S and Fe abundance free and fixed all the other abundances to solar to fit the data.

\begin{figure}
\centering
\includegraphics[width=8cm, angle=270]{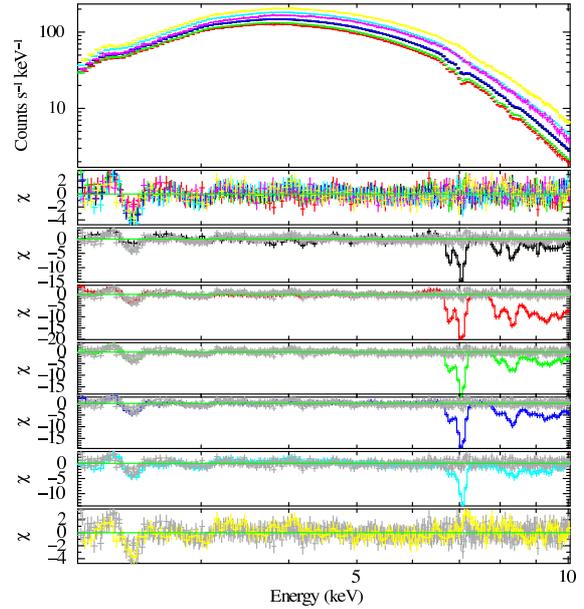}
\caption{\label{fig:3}X-ray spectra of all the six {\it XMM-Newton} observations of 4U~1636$-$47 fitted simultaneously. 
The second panel from the top shows the residuals of the best-fitting model to all 
observations; the following panels are
the residuals of each observation when the strength of the emission and absorption components are set to zero.
Each color corresponds to one observation in Table~\ref{tab:1}, in the sequence 
black, red, green, blue, light blue, magenta and yellow.
We do not show a residual panel for ObsID 0670673101 (magenta) because there are 
no emission or absorption components in this observation.}

\end{figure}

Our best-fitting model contains a moderately broad Gaussian line at 7~keV, consistent with the Ly$\alpha$ line of Fe~{\sc xxvi}. A marginal detection of a similar line, 
likely due to reflection off the accretion disc, had been previously reported in this source \citep{Tomsick2000,Tomsick2014}. 
Since Fe reflection lines should appear between 6.4 keV (Fe {\sc i}) and 6.97 keV (Fe {\sc xxvi}), we constrained the line to be in the range $6.4-7$ keV during the fits. 
The best-fitting value of the energy of the line pegged at the upper limit of this range, 
which could be in partly due to an imperfect calibration of the energy scale in burst-mode.
The reduced $\chi^2$ of our model is $\chi_{\nu}^{2}=0.86$ for 259 d.o.f.
We plot the data and our best-fitting model in Figure~\ref{fig:2}, and show the best-fitting parameters of these two observations using our model in Table~\ref{tab:2}.
The residual panels in this figure show the effect of the different parameters to the fit.
The fit does not require any relativistically Doppler red- or blueshifted emission line.
Instead, the abundance of S in {\sc vphabs} is higher than solar, 1.32$\pm0.06$, and 
that of Fe is lower than solar, 0.54$\pm0.07$.

As a final check, we also created calibrated event files without applying the {\it epfast} correction and fitted the spectra with our model with variable sulphur and iron abundances. Since the best-fitting parameters in this case are consistent, within errors, with those from the fits to the spectra for which we did apply the {\it epfast} correction, we do not show a plot of this analysis. The only difference between the two sets of spectra is that in the model for the data for which we do not apply {\it epfast} we need to add an extra, marginally significant, emission line at 6.65$\pm$0.09~keV in the model of ObsID 0670673101.

\subsection{Fits to the burst- and timing-mode observations using the old calibration}
\begin{figure}
\centering
\includegraphics[width=8cm, angle=270]{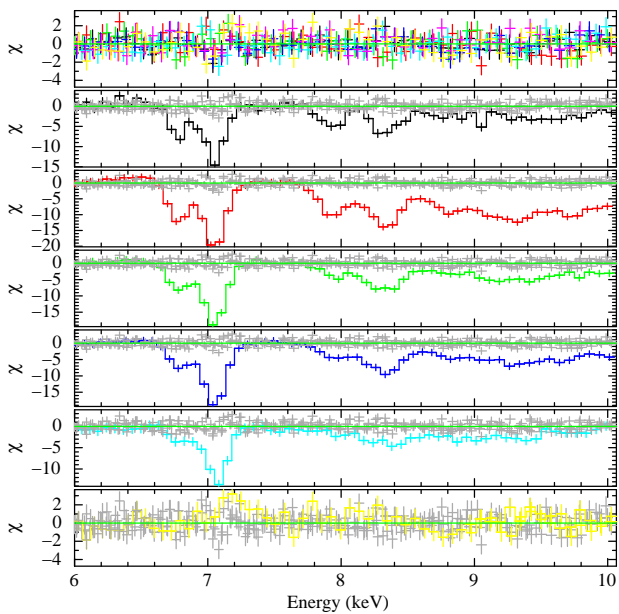}
\caption{\label{fig:4} All the residuals of the burst- and timing-mode observations except ObsID 0670673101.}
\end{figure}

We subsequently fitted our new model to all seven spectra simultaneously.
A quick inspection of the residual plots indicated, in some cases, the presence of absorption features at energies of $\sim$
6.5~keV or higher. Therefore we added up to four Gaussian absorption lines, using negative {\sc gauss} in XSPEC,  
and two edges, {\sc edge} in XSPEC, to account for possible absorption from highly ionised material close to the source. Not all these components were required in all observations.
To keep the model as simple as possible, when the best-fitting parameters of these absorption components turned to be similar within errors, we linked these parameters  
across the observations.
The model we fitted, {\sc vphabs*(diskbb+gauss+powerlaw - gauss$_{1}$ - gauss$_{2}$ - gauss$_{3}$ - gauss$_{4}$)*edge$_{1}$*edge$_{2}$}, 
gives an acceptable fit, with $\chi_\nu^2$ = 0.99 for 895 d.o.f..
We show the best-fitting parameters in Table~\ref{tab:3} and plot the spectra and
best-fitting model of all the observations in Figure~\ref{fig:3}.
In order to show the emission and absorption lines and edges in each observation, we set the strength of these
components to zero in the residual panels (see Figure~\ref{fig:3}). In Figure~\ref{fig:4} we show a zoom in of the residual panels of Figure~\ref{fig:3} in the energy range 6-10~keV.

Compared to the parameters in \cite{Kubota2007}, \cite{Rska2014}
and \cite{DT2014}, 
we find a higher value of $N\rm{_{H}}$ than theirs and
the temperature of the disc is higher than that of \cite{Kubota2007}.
The energy of the absorption lines and edges are consistent with those of Fe~{\sc xxv}~He$\alpha$ (6.70~keV),
Fe~{\sc xxvi}~Ly$\alpha$ (6.97~keV), Ni~{\sc xxviii}~Ly$\alpha$ (8.09~keV) or Fe~{\sc xxv}~Ly$\beta$ (7.88~keV),
Fe~{\sc xxvi}~Ly$\beta$ (8.25~keV), Fe~{\sc xxv}~K-edge (8.83~keV) and Fe~{\sc xxvi}~K-edge (9.28~keV),
similar to the identification in \cite{DT2014}. However, the results are not identical; e.g., \cite{DT2014} reported an absorption edge
at 8.67~keV in ObsID 0670671301 that is not required in our fits. 
\cite{Rska2014} detected seven iron absorption lines with {\it Suzaku}, 
four of which are Fe~{\sc xxv}~He$\alpha$, Fe~{\sc xxvi}~Ly$\alpha$, Fe~{\sc xxv}~Ly$\beta$ and Fe~{\sc xxvi}~Ly$\beta$,
the same ones we report here. The remaining two absorption lines identified by \cite{Rska2014} are Fe~{\sc xxv}~Ly$\gamma$ 
and Fe~{\sc xxvi}~Ly$\gamma$, which we do not detect here.

\begin{table*}
\caption{\label{tab:3}Parameters of the emission and absorption lines and edges in the {\it XMM-Newton} observations of 4U 1630$-$47. }
\renewcommand{\arraystretch}{1.3}
\scriptsize
\setlength{\tabcolsep}{12pt}
\begin{tabular}{ccccccccccccccccccccccccccccccccccccc}
\hline \hline 
\multicolumn{2}{c}{ObsID}&&0670671501-1&0670671501-2&0670671301&0670672901&0670673001&0670673101&0670673201\\
\hline
\multicolumn{2}{c}{$^{a}N\rm{_{H}}$ ($\rm{10^{22}~cm^{-2}}$)}&&14.1$\pm0.1~^{[1]}$&14.1$\pm0.1~^{[1]}$&14.1$\pm0.1~^{[1]}$&14.1$\pm0.1~^{[1]}$&14.1$\pm0.1~^{[1]}$&14.1$\pm0.1~^{[1]}$&14.1$\pm0.1~^{[1]}$\\ 
\multicolumn{2}{c}{$^{b}$S/S$_{\odot}$}&&$1.47\pm0.02~^{[2]}$&$1.47\pm0.02~^{[2]}$&$1.47\pm0.02~^{[2]}$&$1.47\pm0.02~^{[2]}$&$1.47\pm0.02~^{[2]}$&$1.47\pm0.02~^{[2]}$&$1.47\pm0.02~^{[2]}$\\
\multicolumn{2}{c}{$^{b}$Fe/Fe$_{\odot}$}&&$0.95\pm0.04~^{[3]}$&$0.95\pm0.04~^{[3]}$&$0.95\pm0.04~^{[3]}$&$0.95\pm0.04~^{[3]}$&$0.95\pm0.04~^{[3]}$&$0.95\pm0.04~^{[3]}$&$0.95\pm0.04~^{[3]}$\\ 
\hline

\multirow{4}{*}{\multirow{1}{*}{{\sc gauss$_{e}$}}}&E (keV) &&6.89$\pm0.02~^{[4]}$&6.89$\pm0.0~2^{[4]}$&6.89$\pm0.02~^{[4]}$&6.89$\pm0.02~^{[4]}$&6.89$\pm0.02~^{[4]}$&6.89$\pm0.02~^{[4]}$&6.89$\pm0.02~^{[4]}$\\
&$\sigma$ (eV)&&169.1$_{-23.1}^{+13.5}~^{[5]}$&169.1$_{-23.1}^{+13.5}~^{[5]}$&169.1$_{-23.1}^{+13.5}~^{[5]}$&169.1$_{-23.1}^{+13.5}~^{[5]}$&169.1$_{-23.1}^{+13.5}~^{[5]}$&169.1$_{-23.1}^{+13.5}~^{[5]}$&169.1$_{-23.1}^{+13.5}~^{[5]}$\\
&$k_{gau}$ & &4.6$_{-0.5}^{+1.0}~^{[6]}$&4.6$_{-0.5}^{+1.0}~^{[6]}$&4.6$_{-0.5}^{+1.0}~^{[6]}$&4.6$_{-0.5}^{+1.0}~^{[6]}$ &1.3$\pm0.5~^{[7]}$&$<0.9$&1.3$\pm0.5~^{[7]}$\\
&W (eV)&&113.4$_{-53.4}^{+57.5}$&102.0$_{-52.8}^{+51.1}$&84.1$_{-39.8}^{+40.6}$&81.6$_{-39.8}^{+36.4}$&14.9$\pm5.4$&$<12.4$ &11.5$\pm4.4$ \\
\hline

\multirow{4}{*}{\multirow{1}{*}{{\sc gauss$_{a1}$}}}&E (keV)&&6.78$\pm0.01~^{[8]}$&6.78$\pm0.01~^{[8]}$&6.78$\pm0.01~^{[8]}$&6.78$\pm0.01~^{[8]}$&6.78$\pm0.01~^{[8]}$&6.78$\pm0.01~^{[8]}$&6.78$\pm0.01~^{[8]}$\\
 &$\sigma$ (eV)&&10.0$_{-10.0^p}^{+12.2}~^{[9]}$&10.0$_{-10.0^p}^{+12.2}~^{[9]}$&10.0$_{-10.0^p}^{+12.2}~^{[9]}$&10.0$_{-10.0^p}^{+12.2}~^{[9]}$&10.0$_{-10.0^p}^{+12.2}~^{[9]}$&10.0$_{-10.0^p}^{+12.2}~^{[9]}$&10.0$_{-10.0^p}^{+12.2}~^{[9]}$\\
&$k_{gau}$&&3.4$\pm0.5~^{[10]}$&3.4$\pm0.5~^{[10]}$&3.4$\pm0.5~^{[10]}$&3.4$\pm0.5~^{[10]}$&1.2$\pm0.3$&$<0.3$&$<1.1$\\ 
&W (eV)&&65.5$_{-24.2}^{+25.9}$&60.3$_{-19.0}^{+23.2}$&51.7$_{-17.6}^{+18.5}$&50.4$_{-17.5}^{+18.9}$&21.2$\pm7.2$&$<140.9$&$<20.9$\\
\hline

\multirow{4}{*}{\multirow{1}{*}{{\sc gauss$_{a2}$}}}&E (keV)&&7.03$\pm0.01~^{[11]}$&7.03$\pm0.01~^{[11]}$&7.03$\pm0.01~^{[11]}$&7.03$\pm0.01~^{[11]}$&7.03$\pm0.01~^{[11]}$&7.03$\pm0.01~^{[11]}$&7.03$\pm0.01~^{[11]}$\\
&$\sigma$ (eV)&&10.0$_{-10.0^p}^{+4.3}~^{[12]}$&10.0$_{-10.0^p}^{+4.3}~^{[12]}$&10.0$_{-10.0^p}^{+4.3}~^{[12]}$&10.0$_{-10.0^p}^{+4.3}~^{[12]}$&10.0$_{-10.0^p}^{+4.3}~^{[12]}$&10.0$_{-10.0^p}^{+4.3}~^{[12]}$&10.0$_{-10.0^p}^{+4.3}~^{[12]}$\\
&$k_{gau}$&&4.4$\pm0.6~^{[13]}$&4.4$\pm0.6~^{[13]}$&4.4$\pm0.6~^{[13]}$&4.4$\pm0.6~^{[13]}$&3.1$\pm0.3$&$<0.3$&$<0.5$\\
&W (eV)&&95.2$_{-9.5}^{+17.4}$&87.5$_{-10.7}^{+13.9}$&74.9$_{-8.9}^{+14.6}$&73.0$_{-9.9}^{+12.5}$&45.2$_{-3.3}^{+4.4}$ &$<12.0$&$<10.3$ \\
\hline

\multirow{4}{*}{\multirow{1}{*}{{\sc gauss$_{a3}$}}} &E (keV)&&7.93$\pm0.01~^{[14]}$&7.93$\pm0.01~^{[14]}$&7.93$\pm0.01~^{[14]}$&7.93$\pm0.01~^{[14]}$&7.93$\pm0.01~^{[14]}$&7.93$\pm0.01~^{[14]}$&7.93$\pm0.01~^{[14]}$\\
&$\sigma$ (eV)&&10.0$_{-10.0^p}^{+27.6}~^{[15]}$&10.0$_{-10.0^p}^{+27.6}~^{[15]}$&10.0$_{-10.0^p}^{+27.6}~^{[15]}$&10.0$_{-10.0^p}^{+27.6}~^{[15]}$&10.0$_{-10.0^p}^{+27.6}~^{[15]}$&10.0$_{-10.0^p}^{+27.6}~^{[15]}$&10.0$_{-10.0^p}^{+27.6}~^{[15]}$\\
&$k_{gau}$&&0.5$\pm0.06~^{[16]}$&0.8$\pm0.1$&1.0$\pm0.07$&0.5$\pm0.06~^{[16]}$&0.3$\pm0.1$&$<0.2$&$<0.4$\\ 
&W (eV)&&29.2$\pm4.5$&30.2$_{-3.2}^{+2.7}$&13.6$\pm2.1$&13.2$\pm2.1$&6.0$_{-2.1}^{+2.5}$&$<249.0$&$<146.3$\\
\hline

\multirow{4}{*}{\multirow{1}{*}{{\sc gauss$_{a4}$}}} &E (keV)&&8.32$\pm0.01~^{[17]}$&8.32$\pm0.01~^{[17]}$&8.32$\pm0.01~^{[17]}$&8.32$\pm0.01~^{[17]}$&8.32$\pm0.01~^{[17]}$&8.32$\pm0.01~^{[17]}$&8.32$\pm0.01~^{[17]}$\\
 &$\sigma$ (eV)&&67.6$_{-12.7}^{+15.2}~^{[18]}$&67.6$_{-12.7}^{+15.2}~^{[18]}$&67.6$_{-12.7}^{+15.2}~^{[18]}$&67.6$_{-12.7}^{+15.2}~^{[18]}$&67.6$_{-12.7}^{+15.2}~^{[18]}$&67.6$_{-12.7}^{+15.2}~^{[18]}$&67.6$_{-12.7}^{+15.2}~^{[18]}$\\
 &$k_{gau}$&&1.2$\pm0.1~^{[19]}$&1.2$\pm0.1~^{[19]}$&1.2$\pm0.1~^{[19]}$&1.2$\pm0.1~^{[19]}$&0.8$\pm0.1$&$<0.2$&$<0.5$\\ 
&W (eV)&&52.3$_{-2.4}^{+3.5}$&47.1$_{-2.3}^{+3.3}$&38.0$\pm2.4$&36.9$\pm2.4$&18.1$\pm2.4$&$<287.6$&$<182.2$ \\
\hline

\multirow{2}{*}{\multirow{1}{*}{{\sc edge$_{1}$}}} &E (keV)&&8.63$\pm0.02~^{[20]}$&8.63$\pm0.02~^{[20]}$&8.63$\pm0.02~^{[20]}$&8.63$\pm0.02~^{[20]}$&8.63$\pm0.02~^{[20]}$&8.63$\pm0.02~^{[20]}$&8.63$\pm0.02~^{[20]}$\\
&\centering$\tau$~~~~~~~~&&0.12$\pm0.01~^{[21]}$&0.12$\pm0.01~^{[21]}$&0.04$\pm0.01~^{[21]}$&0.04$\pm0.01~^{[21]}$&0.04$\pm0.01~^{[21]}$&$<0.03$&$<0.01$\\ 
\hline

\multirow{2}{*}{\multirow{1}{*}{{\sc edge$_{2}$}}} &E (keV)&&9.05$\pm0.04~^{[22]}$&9.05$\pm0.04~^{[22]}$&9.05$\pm0.04~^{[22]}$&9.05$\pm0.04~^{[22]}$&9.05$\pm0.04~^{[22]}$&9.05$\pm0.04~^{[22]}$&9.05$\pm0.04~^{[22]}$\\
&\centering$\tau$~~~~~~~~&&$<0.04$&0.04$\pm0.01~^{[23]}$&0.04$\pm0.01~^{[23]}$&0.04$\pm0.01~^{[23]}$&$<0.004$&$<0.02$&$<0.02$\\ 
\hline
$\chi^{2}_{\nu}$& \multicolumn{7}{c}{0.99 for 895 d.o.f.} \\ 
\hline
\end{tabular}

\begin{flushleft} 
\begin{tablenotes}
{\bf Notes.}  The {\sc guass$_{e}$} and {\sc guass$_{a1}$} to {\sc guass$_{a4}$} components represent the emission and absorption lines of 
Fe~{\sc xxv}~He$\alpha$,
Fe~{\sc xxvi}~Ly$\alpha$, Ni~{\sc xxvii}~Ly$\alpha$ or Fe~{\sc xxv}~He$\beta$ and Fe~{\sc xxvi}~Ly$\beta$, respectively.\\
The {\sc edge$_1$} and {\sc edge$_2$} components indicate the absorption K-edges of Fe~{\sc xxv} and Fe~{\sc xxvi}, respectively.\\
See Table~\ref{tab:2} for the definition and units of the parameters. The symbols used in this Table have the same meaning as in Table~\ref{tab:2}.  As in Table~\ref{tab:2}, superscripts indicate parameters that were linked between observations during the fits.
\end{tablenotes}

\end{flushleft}

\end{table*}

\begin{figure}
\centering
\includegraphics[width=8cm]{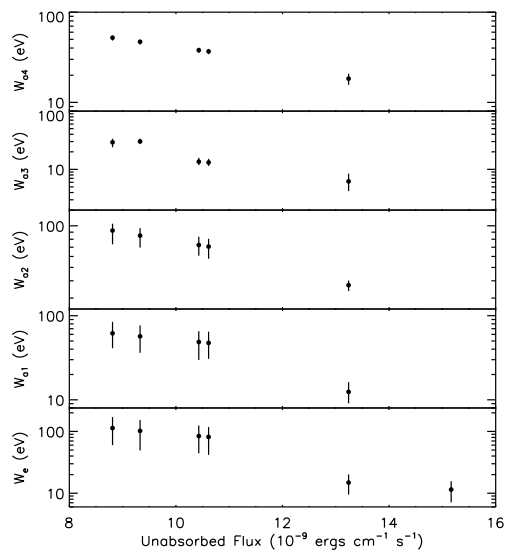}
\caption{\label{fig:5}The equivalent width of the broad emission and narrow absorption lines of 4U~1630$-$47 as a function of the
total unabsorbed flux in the 2-10~keV range. $W_{e}$ and $W_{a_1}$ to $W_{a_4}$ represent, respectively, the equivalent width of the emission and absorption lines of Fe~{\sc xxv}~He$\alpha$,
Fe~{\sc xxvi}~Ly$\alpha$, Ni~{\sc xxvii}~Ly$\alpha$ or Fe~{\sc xxv}~He$\beta$ and Fe~{\sc xxvi}~Ly$\beta$.
}
\end{figure}

We plot the equivalent width of the emission and absorption lines as a function of the total unabsorbed flux in the $2-10$~keV range in Figure~\ref{fig:5}.
From this figure it appears that the equivalent width of the
emission and absorption lines is anti-correlated with the total unabsorbed flux.
To test this we fitted both a constant and a linear function to each of these relations to asses whether the decreasing trend is significant.
For the case of the absorption lines the F-test probabilities range from $3\times10^{-6}$ to $7\times10^{-2}$, indicating that in most cases the decrease of the equivalent width with flux is significant. For the emission line, however, the F-test suggests that a linear function is not significantly better than a constant.
Since the emission and absorption lines in this part of the spectrum respond mostly to the high-energy flux, we also examined the plots of the line equivalent widths vs. the $7-10$~keV flux; the trends are the same as those in Figure~\ref{fig:5} for which we used the $2-10$~keV flux, and hence we do not show those plots here.

\begin{figure}
\centering
\includegraphics[width=8cm]{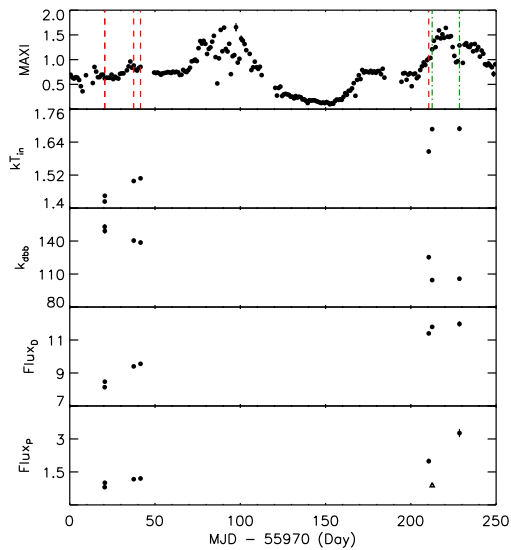}
\caption{\label{fig:6}Light curve and time history of some fitting parameters of 4U~1630$-$47. 
The upper panel shows the MAXI light curve (in units of 
photons $\rm{cm^{-2} s^{-1}}$) in the 2-10~keV band. 
The green dashed-dotted and red dashed vertical lines indicate the times of the two
burst-mode and the four timing-mode {\it XMM-Newton} observations, respectively. 
The second and third panels show, respectively, 
the temperature (in units of keV) and normalisation of the {\sc diskbb} component. 
The fourth and fifth panels show 
the 2-10~keV unabsorbed flux, in units of 
$\rm{10^{-9}}$~erg cm$\rm{^{-2}}$ s$\rm{^{-1}}$, of the {\sc diskbb} and {\sc powerlaw} components, respectively. 
The triangle in the fifth panel denotes the 95\% confidence upper limit of the flux
of the {\sc powerlaw} component in that observation. Some of the error bars are too small to show up on this plot.
}
\end{figure}

In Figure~\ref{fig:6} we plot some of the fitting parameters as a function of time.
In the top panel of Figure~\ref{fig:6} we show the
MAXI (Monitor of All-sky X-ray Image, \citealt{Matsuoka2009}) light curve of 4U~1630$-$47 in 
the 2-10~keV energy band from MJD 55970 to MJD 56220.
The other panels show, respectively, the time histories of the temperature, the normalisation and 
the flux of the {\sc diskbb} component, and the flux of the {\sc powerlaw} component. 
We do not plot the emission {\sc gauss} component in this figure
because we linked the parameters of this component across several observations (see Table~\ref{tab:3}).
Figure~\ref{fig:6} shows that the temperature of the {\sc diskbb} component generally increases with time, 
whereas the normalisation shows the opposite trend;
on average, the flux of the {\sc diskbb} and {\sc powerlaw} components appear to increase with time. 
The temperature of the disc and the disc and power-law fluxes are generally correlated, 
whereas the disc normalisation is anti-correlated, with the 2-10 keV MAXI flux. 
This is consistent with the standard scenario of black-hole states, but given the long time gaps between the observations, 
and the complex changes of the light curve with time (top panel of Figure~\ref{fig:6}), we do not discuss these correlations further.   
The {\sc powerlaw} component in ObsID 0670673101 is not significant, 
and therefore we plotted the upper limit as a triangle.
From this figure it is apparent that the emission in the 2-10~keV range is always dominated by the {\sc diskbb} component.

\subsection{Fits to two burst-mode observations using the new calibration}
\label{Sec:new}
While we were analysing these data, the {\it XMM-Newton} team released a new set of calibration files (dated March 31 2015) 
for Epic-pn burst-mode observations. We therefore extracted the burst-mode spectra again using the new calibration, and 
fitted our model to these two observations of 4U~1630$-$47. 
Comparing to the previous fitting results of the two burst-mode observations using the old calibration,
we added an extra negative {\sc gauss} component to the model to account for a possible absorption line at $\sim7$ keV. 
We list our best-fitting parameters in Table~\ref{tab:4}.
The F-test probabilities for the emission and absorption lines are, respectively, $10^{-2}$ and $10^{-3}$, indicating that 
the lines are marginally significantly.
We also tried to fit these data using the model of \cite{DT2013}, but the three emission lines are not significantly detected.
Finally, we also reduced these two burst-mode observations using the new calibration without applying the {\it epfast} correction. 
The best-fitting parameters are consistent, within errors, with those from the other fits, and in this case we do not find any significant emission line in ObsID 0670673201 either.

\begin{table}
\caption{\label{tab:4}Best-fitting parameters for the two burst-mode observations of 4U~1630$-$47 based on the new calibration files using our model.}
\renewcommand{\arraystretch}{1.3}
\scriptsize
\centering
\begin{tabular}{llcc}
\hline \hline
ObsID &&0670673101 &0670673201 \\
\hline

{\sc tbabs/vhpabs} & $N\rm{_{H}}$ ($\rm{10^{22}~cm^{-2}}$) & $13.8\pm0.2~^{[1]}$ & $13.8\pm0.2~^{[1]}$\\
& S/S$_{\odot}$ & $1.28\pm0.06~^{[2]}$ & $1.28\pm0.06~^{[2]}$\\
& Fe/Fe$_{\odot}$ & $0.74\pm0.08~^{[3]}$ & $0.74\pm0.08~^{[3]}$\\
\hline

\multirow{2}{*}{{\sc diskbb}}&Tin (keV) & 1.65$\pm0.02$ & 1.66$\pm0.02$\\
&$k_{dbb}$ & 115.2$\pm3.9$ & 122.5$\pm4.5$\\
\hline

\multirow{2}{*}{{\sc powerlaw}}&$\Gamma$ & $2^f$ & $2^f$\\
&$k_{pow}$ & $0.26\pm0.09$& $1.22\pm0.1$\\
\hline

\multirow{4}{*}{{\sc gauss$_{e}$}}&E (keV) & $6.58_{-0.09}^{+0.38}~^{[4]}$ & $6.58_{-0.09}^{+0.38}~^{[4]}$\\
&$\sigma$ (eV)& $135.7_{-74.6}^{+181.3}~^{[5]}$ &$135.7_{-74.6}^{+181.3}~^{[5]}$\\
&$k_{gau}$&1.4$_{-0.5}^{+92.1}$ &$<2.1$\\
&W (eV)& 13.9$_{-13.9}^{+15}$&$<32.4$\\
\hline

\multirow{4}{*}{{\sc gauss$_{a}$}}&E (keV) & $6.99_{-6.99p}^{+0.06}~^{[6]}$ & $6.99_{-6.99p}^{+0.06}~^{[6]}$\\
& $\sigma$ (eV)& 87$_{-86.4}^{+159}~^{[7]}$ & 87$_{-86.4}^{+159}~^{[7]}$ \\
&$k_{gau}$&1.3$_{-20.4}^{+0.5}$ &$<1.6$\\
&W (eV)& 16.2$_{-15.6}^{+11.2}$&$<9.2$\\
\hline

$\chi^{2}_{\nu}$& \multicolumn{2}{c}{0.74 for 256 d.o.f.}\\
\hline

\end{tabular}
\begin{flushleft}
{\bf Notes.} See Table~\ref{tab:2} for the definition and units of the parameters. 
The symbols used in this Table have the same meaning as in Tables~\ref{tab:2} and \ref{tab:3}.
 Also, as in Table~\ref{tab:2}, superscripts indicate parameters that were linked between observations during the fits. 
\end{flushleft}
\end{table}

\section{DISCUSSION}
Recently, \cite{DT2013} reported the detection of three Doppler-shifted emission lines 
arising from the jet of 4U~1630$-$47 in an XMM-Newton observation obtained during the 2012 outburst of the source.
Here we show that this same observation can be well fitted with a model that does not require the three emission lines. 
The main difference between our model and that of \cite{DT2013} is that we allow the abundance of S and Fe in the interstellar material along the line of sight to the source to vary. 
Our model also fits well the other observation in \cite{DT2013}, in which they do not detect  the emission lines. Fitting these two observations simultaneously, 
we find that the abundances of S and Fe in the interstellar medium toward the source are, respectively, 1.32$\pm0.06$ and 0.54$\pm0.07$, in solar units.
Because of the large value of the column density in the interstellar medium toward the source, 
a non-solar abundance of these elements impacts upon the model at energies around the neutral Fe edge, 
at $\sim7.1$ keV (see the two bottom panels in Figure \ref{fig:2}), such that the emission lines are no longer required in the model. 
Our model also fits the rest of the XMM-Newton observation of the 2012 outburst of 4U~1630$-$47 \citep{DT2014}; 
in this case, similar to \cite{DT2014}, we need to add several absorption lines and edges in the $6.7-9.1$~keV energy range, likely due to photo-ionised material close to the source. 

Since the two models are fundamentally different,  we cannot compare them from a statistical point of view (e.g., using the F-test); 
however, our model fits the same data with less free parameters, and it is therefore simpler than the one of \cite{DT2013}. 
Our model does include a moderately broad ($\sigma=183_{-79}^{+108}$~eV) emission line at $7$ keV. 
This line is consistent with a marginally significant line  detected from this \citep{Tomsick2000,Tomsick2014,King2014} and other sources \citep[e.g.][]{Miller2007}, 
which is usually interpreted as due to emission from the hard (power-law) component reflected off the accretion disc, 
with the broadening being due to relativistic effects close to the black hole \citep[e.g.][]{Fabian2012}. 
The fact that in our fits the best-fitting energy of this line pegs at the upper limit, $7$ keV, 
that we imposed in the model may be partly due to the uncertainties in the energy calibration of the Epic-pn burst mode (see below), 
or to inaccuracies in the cross section tables in the component that we used to fit the interstellar absorption. 

Our model yields non-solar abundances of S and Fe in the ISM toward 4U 1630$-$47. In the case of Fe, our best-fitting abundance is consistent with measurements 
of the Fe abundance in the ISM toward nine low-mass X-ray binaries (not including 4U~1630$-$47) using high-resolution spectra from the Reflection Grating 
Spectrometer on board {\it XMM-Newton} \citep[][these authors did not measure the S abundance]{Pinto2013}. 
This makes our model plausible and, if correct, this could explain why the jet lines were not observed with other satellites \citep{Hori2014,Neilsen2014}.

While we were analysing these data, the {\it XMM-Newton} team released a new set of calibration files for the Epic-pn burst-mode observations. 
Using this new calibration, we do not detect the lines reported by \cite{DT2013} either, not even if, as they did, we assume solar abundances for all elements in the ISM (see \ref{Sec:new}). 
In this case the sulphur and iron abundances, respectively S$/$S$_\odot=1.28\pm0.06$ and Fe$/$Fe$_\odot=0.74\pm0.08$, are consistent within errors with the ones that we obtained using the old calibration files.
Assuming that the new calibration is better than the previous one, this result casts doubt on the presence of the Doppler-shifted lines in this source and, at the same time, 
it provides a rough estimate of the relative accuracy of the calibration of the Epic-pn burst-mode data used by \cite{DT2013}.

The model that we propose here provides also a good fit to all six {\it XMM-Newton} observations of this source during the 2012 outburst. 
Similar to \cite{DT2014}, we find a number of absorption lines and edges in the spectra of the 
four {\it XMM-Newton} observations that were obtained using the EPIC-pn camera in timing mode.
These lines and edges are consistent with being due to Fe~{\sc xxv}, Fe~{\sc xxvi} and Ni~{\sc xxviii}, 
indicating the existence of highly ionised material in the vicinity of the source. 
When we fit all observations simultaneously, our model yields a significantly higher Fe abundance, $0.95\pm0.04$, 
than in the case when we fit only the two burst-mode observations (the S abundance is consistent with the one we obtained from the two burst-mode observations). 
This could be due to the fact that we used individual Gaussian lines and edges, instead of a self-consistent model of a warm absorber \cite[see, e.g.,][]{DT2014}, to fit the absorption by this highly ionised material.
In order to keep our model as simple as possible, we linked the parameters of the absorption lines and edges across the different observations whenever possible. 
This prevents us from carrying out a detailed analysis of the absorption features in the observations separately. 
We note, however, that a recent study of 4U~1630$-$47 with high spectral resolution data from {\it Chandra} \citep{Miller2015}, showed evidence of at least two separate absorption zones in the disc wind component in this source.
Since it is not the purpose of this paper to discuss 
the disc wind in this sourse in detail (this aspect of the {\it XMM-Newton} data presented here was already discussed by \cite{DT2014}), we did not explore this possibility further.

Finally, we find that the equivalent width of the emission and absorption lines
is anti-correlated with the $2-10$~keV unabsorbed flux.
For the absorption lines, the change of the equivalent width with flux is probably due to changes in the ionisation fraction of the ionised material. 
For the emission line, the drop of the equivalent width could be interpreted as either a change of the ionisation fraction \citep{Garc},
or the effect of light bending close to the black hole \citep{Miniutti}. 
If the ionised material that produces the absorption lines and edges is part of the accretion disc \citep[e.g., the disc atmosphere,][]{Rska2014}, 
this same material would be the one that produces the reflection component and hence the moderately broad Fe emission line. 
In that case a single mechanism, a change in the ionisation fraction of the disc, would be responsible for the drop of the equivalent width of both 
the absorption and emission lines as the flux increases.

\section*{ACKNOWLEDGEMENTS}

This work has made use of data from the High Energy Astrophysics Science Archive Research Center (HEASARC), provided by NASA/Goddard 
Space Flight Center (GSFC). This work is partly supported by China Scholarship Council (CSC), under the grant number 201404910530.
We thank an anonymous referee for his/her constructive comments that helped us improve this paper.

\bibliographystyle{mnras}
\bibliography{kapteyn_2014.bib}
\end{document}